\documentclass[12pt,preprint]{aastex}
\shorttitle{Kilogauss magnetic fields in white dwarfs}
\shortauthors{Valyavin et al.}
\begin{document}

\title{A search for kilogauss magnetic fields in white dwarfs and
hot subdwarf stars}

\author{G. Valyavin\altaffilmark{1,2}, S. Bagnulo\altaffilmark{3},
S.~Fabrika\altaffilmark{2}, A. Reisenegger\altaffilmark{4},
G.A.~Wade\altaffilmark{5}, Inwoo Han\altaffilmark{1},
D. Monin\altaffilmark{6}}

\altaffiltext{1}{Korea Astronomy and Space Science Institute, 61-1,
Whaam-Dong, Youseong-Gu, Taejeon, Republic of Korea 305-348 }
\altaffiltext{2}{Special Astrophysical Observatory, Russian Academy
of Sciences, Nizhnii Arkhyz, Karachai Cherkess Republic, 357147,
Russia}
\altaffiltext{3}{European Southern Observatory, Alonso de
C\'ordova 3107, Santiago, Chile}
\altaffiltext{4}{Departamento de Astronom{\'\i}a y
Astrof{\'\i}sica, Pontificia Universidad Cat\'olica de Chile,
Casilla 306, Santiago 22}
\altaffiltext{5}{Physics Department, Royal
Military College of Canada, Kingston, Ontario, Canada}
\altaffiltext{6}{Dominion Astrophysical
Observatory, 5071, West Saanich Road, Victoria, British Columbia, V9E 2E7,
Canada}

\begin{abstract}
We present new results of a survey for weak magnetic fields among DA
white dwarfs with inclusion of some brighter hot subdwarf stars.
We have detected variable circular polarization in the
H$\alpha$ line of the hot subdwarf star Feige~34 (SP: sdO).
From these data, we estimate that the longitudinal magnetic field of
this star varies from $-1.1 \pm 3.2$\,kG to $+9.6 \pm 2.6$\,kG, with a
mean of about $+5$\,kG and a period longer than 2~h. In this
study, we also confirm the magnetic nature of white dwarf WD1105-048, found
earlier in a study by \citet{AZC04}, and present upper limits of
kilogauss longitudinal magnetic fields of 5 brightest DA white
dwarfs. Our data support the finding of \citet{AZC04} that $\sim$
25\% of white dwarfs have kilogauss magnetic fields. This frequency also
confirms results of early estimates obtained using the magnetic field
function of white dwarfs \citep{FV99}.
\end{abstract}


\section{Introduction}

The investigation of white dwarf stars (WDs) is of fundamental
importance for the understanding of stellar and galactic evolution,
as WDs represent the final evolutionary stage of more than 90\% of
all stars. Nowadays, we believe that the general properties and
evolution of WDs are fairly well understood. However, there are
several important problems that still need to be properly addressed,
especially those connected with the group of about one hundred
isolated magnetic white dwarfs (MWDs)
\citep{ABL81,SS95,LBH03,VBFM03,AZC04}. The first problem is related
to the origin of the MWDs. Earlier studies \citep{ABL81} suggested
that MWDs are descendants of magnetic Ap/Bp stars. More recent
studies, however, have suggested that the progenitors are not
restricted to this class \citep{LBH03,KW04,WF05}. The second problem
is associated with the evolution of a global magnetic field during a
WD's life: it has been shown \citep{LS79,FV99,VF99,LBH03} that the
magnetic field exhibits some peculiar features during the WD life time.
One of them is that WDs with strong fields show a tendency to
increase in fractional incidence with age, in contradiction to the
hypothesis that WD magnetic fields decay with time \citep{WVS87}.

Another group of problems is related to astroseismology of WDs. A
model analysis of pulsation modes applied to pulsating WDs
\citep[for instance]{WNC94}, theoretical exploration \citep{MTV94},
and high-resolution spectroscopy \citep{KDWA98} suggest the general presence
of kilogauss magnetic fields in pulsating white dwarfs. In spite of that,
attempts to measure magnetic fields in pulsating WDs \citep[for
instance]{SG97} have yielded no positive results, suggesting that
the theory of WD pulsation may require revision.

The implications of these studies are extremely significant.
Unfortunately, {\it they are based on a highly biased and limited
sample of strongly magnetic WDs}, and therefore they are still
controversial and require better statistics \citep{LBH03}. It
appears quite likely that the fraction of MWDs (as compared to the
total population of known WDs) will increase significantly if
high-precision spectropolarimetric surveys are conducted.
The required accuracy of magnetic field measurements is
about 1~kG or better \citep{LBH03}.

In order to better
understand the origin of magnetic fields in late type stars,
it is also interesting
to extend observations to the group of hot subdwarfs,
the magnetic nature of which has already been reported \citep{E96,TJFH05},
but which is still not well studied. The importance of a study of hot subdwarf
stars to the theory of stellar evolution
is established by the fact that they exhibit a variety of
evolutionary channels to the white dwarf stage
\citep{GS74,H86,SBKL94,WMG01,WMMG01,M04}.
A search for magnetic
fields among these stars should help to understand the magnetic nature
of some low- normal-mass magnetic white dwarfs which potentially could be
evolutionary
products of the subdwarf stars. In particular, the study of a heterogenous
group of sdO stars \citep[and references therein]{M04}
for which it has been found that about 65\% of them may be unresolved binary systems
\citep{WMMG01}, makes it possible to consider
dynamo-induced magnetic fields in these systems, suggesting a non-fossil field
origin in their magnetic white dwarf decendants.

For these reasons, we are
carrying out an observational program with the 6\,m and 8\,m Russian
and European telescopes. The main goal of the project is the accumulation
of observational data on white dwarf magnetism in the kilogauss region
and statistical analysis of these data for a study of the evolution
of magnetic fields in degenerate stars. Our secondary goals are
monitoring studies of some individual weak-field magnetic WDs
\citep{VBM05} and some individual brighter hot subdwarf stars.
Here, we discuss new results from our survey.
We present the positive detection of a kilogauss longitudinal
magnetic field on the sdO star Feige~34 and confirm the
magnetic nature of white dwarf WD1105-048, detected recently by \citet{AZC04}.
We also suspect the presence of a varying weak longitudinal magnetic
field on the pulsating white dwarf WD1647+591.

\section{Selection of the sample and observational strategy}

At the present stage, our study of hot subdwars is restricted to observations
of only one of the brightest objects Feige~34. Our observations of white
dwarfs are aimed mainly at studying a random sample of WDs in a
limited space volume with an accuracy of magnetic field measurements
of about 1-2\,kG and better. To answer the question of {\it whether
WD magnetic field evolution can be detected in observations}, it is
important to extend field measurements with uniform accuracy to the
whole range of WD masses and temperatures.  All types from the
hottest (youngest) WDs to the coolest (oldest) degenerates of DA8-9
spectral classes are present in our list, in order to provide the
survey with appropriate statistics.

With the aid of the 6\,m telescope (BTA), we searched for circular
polarization in cores of the hydrogen lines of the brightest ($V <
14$\,mag) northern hot (young) WDs and WDs of intermediate
temperatures.  The cooler (and older) WDs, however, can not be
studied well with this telescope. These stars are intrinsically
fainter and have weaker Balmer absorption features that make it
difficult to study them polarimetrically with the necessary field
measurement accuracy. In order to minimize this observational bias,
we have extended our list toward observations with the VLT of a
random sample of southern WDs with $T_{\rm eff} < 9000~K$. These
observations, when completed, will make it possible to determine the
fractional incidence of magnetism in the low-field regime among WDs
of different ages. Our full list consists of 40 isolated WDs of
different masses and temperatures. As follows from \citet{FV99} and
from the discussion presented by \citet{AZC04}, the fractional
incidence of magnetism in the region from 1\,kG to 10\,kG is
expected to be a few to several tens of percents. \citet{AZC04} give
an estimate of 25$\%$. Their conclusion, if true, suggests that weak
magnetic fields may be found in about 10 WDs in our list at a
detection level around 1\,kG. These data, combined with those
obtained for the group of strongly magnetic MWDs, will then be used
to determine unbiased relative fractions of young and old MWDs, to
be compared for a study of the evolution of WD magnetic fields.

In this paper, we report observations of the 6 brightest WDs and
one subdwarf star in our
target list for the 6\,m telescope. Although the survey is not yet
complete, new weak-field magnetic stars are interesting enough to
warrant a separate paper concerning the brightest, hot WDs and WDs
of intermediate temperatures. These data represent our first scientific
observations obtained with a new polarimeter (see next section).
White dwarfs presented here may also be considered as a random sample,
to be compared with
data presented by \citet{AZC04}. The full survey, including the data
from the VLT and a final, detailed statistical analysis will be
presented later.

\section{Observations and data reduction}

The observations were carried out at the 6\,m Russian telescope from
2003 to 2005 in the course of about 9 observing nights shared with
other observational programs. The observations are now obtained
using the updated prime-focus spectrograph-polarimeter UAGS (R~$\sim
2000$, H$\alpha$ region). The instrument and observational technique
are described in detail by \citet{ABVD95} and by \citet{NVF02}.  The
modulation technique we use in observations with this instrument,
the strategy of the observations, and the data reduction are very
similar to those performed by \citet{BSWLM02} and described by
\citet{VBM05}.

In polarimetric observations of each star, we obtain series of
short, consecutive exposures at two orthogonal orientations of the
quarter-wave plate (the sequence of its position angles is
$+45^\circ , -45^\circ , -45^\circ , +45^\circ$). Assuming \textit{a
priori} that the time-scale of possible variability of the
longitudinal magnetic field could be as short as a few minutes, we
usually set the integration time of each exposure to 60-300\,s,
depending on stellar magnitude and sky conditions.

For longitudinal magnetic field measurements, we initially use
cross-correlation analysis of the displacement between positions of
the H$\alpha$ line in spectra of opposite circular polarizations
\citep{MFV02,VBM05}. Applying this method to the series of short
time exposures, we then analyze rows of the longitudinal field
values to rule out a possible variability of the magnetic field on
longer time scales (tens of minutes and longer). After the
determination of these scales for each star, we combine spectra of
equal orientations of the quarter-wave plate into longer equivalent
exposures and build Stokes~$I$ and $V$ profiles as described in
\citet{VBM05}.

Finally, we obtain the longitudinal field determinations from the
Stokes~$I$ and $V$ profiles through the weak-field approximation, as
described by \citet{BSWLM02}. Associated error bars are obtained by
using the Monte Carlo modeling method demonstrated by \citet{SS94}.

\section{Results}

Results of longitudinal magnetic field measurements are summarized
in Table~1, where column~1 is the name of the WD, column~2 is the
spectral class, column~3 is the Julian Date of the midpoint of the
observation, column~4 is the exposure time, and columns 5-6 report
the measurements and uncertainties of the longitudinal magnetic
fields as obtained using the weak-field approximation. The Stokes
$I$ and $V$ spectra are illustrated in Figure~1.

Six targets of our list (WD0009+501, WD0644+375,
WD0713+584, WD1105-048, WD1134+300, WD1647+591) have
already been observed by \citet{SS94,SS95,VBFM03,AZC04,VBM05}. In
order to minimize the probability to observe a possible zero
crossover of the longitudinal magnetic field due to rotation, we
repeated the observations of these WDs.

In order to clarify our polarimetric measurements, here we also
present an example of our observations of the magnetic white dwarf
WD0009+501 \citep{VBM05}. The field of WD0009+501 varies with the
rotation phase from about -120~kG to about +50~kG \citep{VBM05}.
Here, we use its phase-resolved Stokes $I$ and $V$ spectra at the
maximum field obtained during JD245300(3-8). As one can see, the
comparatively weak magnetic field (~50~kG) of this faint (for
spectropolarimetry) WD ($V=14.4$) can easily be resolved and
measured in our observations.

Two targets in our list WD1105-048 and Feige~34 showed
longitudinal magnetic fields by the presence of weak circular polarization
in the H$\alpha$ cores of their spectra. We briefly discuss these
results.

{\bf Feige\,34} or {\bf WD1036+433} \citep{MS99} is one of
the brightest ($V = 11.22$) weak-lined subdwarf stars spectroscopically
classified as sdO \citep[and references therein]{GS74,MS99}.
\citet{TMS91,TUM95} suggested this
star to be a close binary system having a companion of spectral type K.
In three observations of this star in different nights,
the longitudinal magnetic field was detected once at more than the
3$\sigma$ level ( see Table~1 ). S-shaped circular polarization is
seen in one of the observations (at JD=2453005.49; see Fig.~1). We
therefore conclude that WD1036+433 may be another one candidate to
magnetic subdwarf stars \citep{E96,TJFH05} with a weak magnetic field.
The peak field is below 10~kG. In
our observations, the field varies in the range from -1~kG to
+9.6~kG, due to possible rotation with a period longer than 2 hours.

{\bf WD1105-048} is, in comparison, a well-studied, ordinary DA3 WD
\citep{MS99} discovered recently as magnetic by \citet{AZC04}. The
longitudinal magnetic field of this star was found to be variable,
varying from about -1 to -4.6~kG. In this study, we confirm the
magnetic nature of this degenerate star. In the course of two
consequent nights, WD1105-048 showed a variable longitudinal field,
from 0 to about -8~kG. From our data, the rotation period is longer
than 3~hours.

In addition, we have selected 2 white dwarfs.

{\bf WD1647+591} is a representative of the group of pulsating WD
stars, the magnetic nature of which has been discussed by a number
of authors \citep[and references therein]{SG97}. Earlier
observations of the longitudinal magnetic field \citep{SG97} yielded
null results at a level of about 3~kG. However, the probability to
observe a possible zero crossover of the field due to rotation of
the star was quite high in those observations. In particular, their
results and conclusions are based on only three measurements with
associated error bars of 1.1, 1.2 and 5.7~kG. These three
observations, obtained in different years, together with the
indirectly estimated rotation period of about 9\,hr \citep[and
references therein]{SG97} do not exclude a situation that the
observations have been obtained close to the crossover. For this
reason, we continued observations of WD1647+591. The data showed no
detection above the 3$\sigma$ level. Just one of the observations
showed a 2.7$\sigma$ detection at JD2453195.46. Unfortunately, poor
weather conditions limited our observations of this star, making it
necessary to continue the monitoring.

{\bf WD\,0501+527} is one of the interesting WDs in our list.
In the high-resolution observations of this star
\citet{RW88} have found a weak emission line at the H$\alpha$
core. \citet{RW88} have argued that the observed emission
can not be due to a nearby red dwarf companion, suggesting a photospheric
origin. Explaining this emission feature they tested an atmosphere model
under the assumption of a chemically stratified atmosphere which can
produce the necessary temperature inversion giving rise to emission cores in the Balmer
lines \citep[and references therein]{RW88}. The presence of magnetic fields
in this object, if detected, could potentially be interesting for
alternative, magnetoionic heating models of the observed emission
\citep[see their discussion]{GM85}. High-resolution spectra
presented by \citet{RW88} make it possible to estimate roughly the surface
magnetic field on WD\,0501+527
to be less than 100~kG. In our observations we estimate its magnetic field
to be weaker than 10~kG.

The magnetic field estimates in the spectra of the other observed
stars did not yield any detections. Unfortunately, due to
weather conditions and technical limitations which require us
to carry out the
observations only in the H$\alpha$ region, we did not achieve the
required accuracy better 2~kG for all the stars. Nevertheless, the
possible detection of the longitudinal magnetic field in WD1036+433
and the confirmation of the magnetic nature of WD1105-048 make it
possible to contribute to the ongoing discussion about the
fractional incidence of WD magnetism in the kilogauss region.

\section{Discussion and conclusion}

We have presented new observations of kilogauss longitudinal
magnetic fields in 6 WDs and one hot subdwarf star. We confirm the magnetic
nature of WD1105-048 and find a new candidate weak-field sdO star,
Feige\,34 (WD1036+433). We did not detect any significant field in the
pulsating WD WD1647+591; nevertheless, we find it important to note
that the 2.7$\sigma$ result (at JD2453195.46), as well as the
systematic negative magnetic field in all of our observations and
those of \citet{SG97} may indicate the presence of a weak, varying,
or fluctuating non-zero longitudinal magnetic field. It should also
be noted that recent high-resolution spectroscopy of WDs
\citep{KDWA98} showed significant broadening of H$\alpha$ cores in
the spectra of pulsating ZZ~Ceti WDs with typical projected
velocities of 30 -- 40 km/s. Such velocities seem to be in conflict
with astroseismology \citep{KDWA98} and support independently the
probable presence of alternative broadening mechanisms such as a
kilogauss magnetic field. For this reason, we hope that polarimetric
observations of WD1647+591 will be continued.

Together with another subdwarf stars the magnetic nature of which
has been recently detected \citep{E96,TJFH05}, the presence of kilogauss
magnetic field on Feige\,34, if confirmed by future observations,
would be of a great interest for theories of the origin of hot subdwarfs.
Despite the fact that Zeeman polarimetric
observations of these stars are relatively uncommon in the
literature, such a high rate of positive detections
suggests that the presence of global magnetic fields may be a typical property
of the hot subdwarfs (by analogy with the Ap/Bp stars).
At this moment, the origin of magnetic fields in these stars is unclear
\citep{TJFH05}. These fields may be fossil remnants of their progenitor fields
or dynamo-generated by any unknown mechanism. In this connection,
the suggested \citep{TMS91,TUM95} binarity of Feige\,34 gives
new interesting possibilities to explain the magnetic nature of
this object. Following the accretion model of helium degenerate
dwarf formation in a close binary \citep{IT86} we may speculate that
the kilogauss magnetic field could be generated by the accretion processes
which might take place in Feige\,34. This assumption, if true,
could then explain the origin of some kilogauss-strength magnetic fields in
low-mass white dwarfs which may potentially be products of
the evolution of the sdO stars.

Kilogauss upper limits are presented for the other 5 WDs. Here, we
formally estimate the fraction of kilogauss MWDs as 17\% (1/6) that
is consistent with the estimate of \citet{AZC04} (25\%).
Note, that practically the same estimate of the frequency of kilogauss MWDs
has been done using the magnetic field function technique \citep{FV99}.
Based on these results, one might speculate that the MWDs are not a
unique class of degenerate stars, but rather represent the
strong-field tail of a continuous distribution of field intensities.
The 10\,\% fraction of megagauss MWDs \citep{LBH03}, in comparison
to 25\,\% for kilogauss fields, supports this idea.
This conclusion, however, could be more definite if framed in terms
of surface magnetic fields (mean field modulus, $B_s$) since our
study is based on observations of only the longitudinal component
$B_l$ of the field, which is smaller than $B_s$ in all cases. In
case of a dipolar geometry, the difference may be as much as 3 times and
higher, depending on the orientation of the dipole to the line of
sight. Furthermore, the large-scale magnetic field structure of the
weak-field MWDs may be different from dipolar, giving a very strong
observational bias to the underestimation of the incidence of WD
magnetism in the polarimetric observations. Therefore, the above
conclusion about the incidence of magnetism below 1\,kG can only be
considered as a lower limit (see also \citet{AZC04}).

Assuming magnetic field conservation during MS star evolution into a
WD, we have to conclude that the majority of MWDs (namely those with
kilogauss field strengths) descend from MS stars with global
magnetic field strengths even weaker than 1~G, much weaker than those of
Ap/Bp stars \citep{P99}, for which dipole fields larger than about
300~G \citep{A04} are found. So, we may assume that stars other than
Ap/Bp stars evolve into kilogauss MWDs \citep{KW04,AZC04,WF05}.
Alternatively, we should conclude that significant flux loss occurs
during the post-main sequence evolutionary process, although this is
unlikely to be consistent with the newly found, high fraction of
MWDs compared to the few percent fraction of Ap/Bp stars on the
upper main sequence. Further examination of these two possibilities
will be among the goals of our more detailed study upon completion
of our survey.

\acknowledgments

We wish to thank Drs. A.\,Pramsky and A.\,Burenkov for their help
during observations. We are also especially grateful to
our anonymous referee and Prof. J.\,Liebert for their valuable comments and
suggestions. GV is grateful to the Korean MOST (Ministry of
Science and Technology, grant M1-022-00-0005) and KOFST (Korean
Federation of Science and Technology Societies) for providing him an
opportunity to work at KAO through Brain Pool program. AR
acknowledges support from FONDECYT Regular Grant 1020840. GAW
acknowledges Discovery Grant support from the Natural Sciences and
Engineering Research Council of Canada.

\clearpage

%%%%%%%%%%%%%%%%%%%%%%%%%%%%TABLE%%%%%%%%%%%%%%%%%%%%%%%%%%%%%%%%%%%%%%%%%%%
\begin{deluxetable}{cccccc}
\tablecolumns{5} \tablewidth{0pc} \tablecaption{Determination of the
mean longitudinal field obtained as explained in \S~3. } \tablehead{
Name & SP& JD& Exp& $B_\mathrm{l}$&
$\sigma$\\
WD    &      & (-2400000)&(sec)&$(kG)$&$(kG)$   \\
}
\startdata
0501+527&DA1      &53004.49&3600&-3.9 & 2.8\\
\hline
0644+375&DA2      &53004.59&4020&+1.9 & 1.8\\
\hline
0713+584&DA4      &53003.53&7100&+0.6 & 0.8\\
     &         &53003.61&7100&+0.8 & 1.0\\
&         &  AVERAGE      &    &+0.7 & 0.6\\
\hline
1036+433&sdO  &53005.49&6000&+9.6 & 2.6\\
     &         &53005.63&5000&+5.9 & 2.8\\
     &         &53422.34&5500&-1.1 & 3.2\\
     &         & AVERAGE & &+5.5& 1.6 \\
\hline
1105-048&DA3      &53006.54&3600&-7.9 & 2.6\\
     &         &53007.55&7200&-4.8 & 2.3\\
     &         &53007.64&3100&+0.1 & 2.7\\
&         &   AVERAGE     &    &-4.4 & 1.4\\
\hline
1134+300&DA3      &53004.64&1800&+8.9& 4.5\\
     &         &53006.62&3600&+3.5 & 2.7\\
&         &   AVERAGE     &    &+4.9 & 2.3\\
\hline
1647+591&DAv4     &53194.47&3600&$-2.9$ & 3.4\\
     &         &53195.46&3600&-8.7 & 3.2\\
     &         &53422.59&5100&-0.5 & 2.8\\
&         &   AVERAGE     &    &-3.8 & 1.7\\
\enddata
\end{deluxetable}

%%%%%%%%%%%%%%%%%%%%%%%%%%%%%%%%%%%%%%%%%%%%%%%%%%%%%%%%%%%%%
%%%%%%%%%%%%%%%%%%%%%%%%%%%%%%%%%%%%%%%%%%%%%%%%%%%%%%%%%%%%%%%%%%%%

\begin{figure}
\centering
\includegraphics[width=4.03cm, height=9.2cm, angle=270]{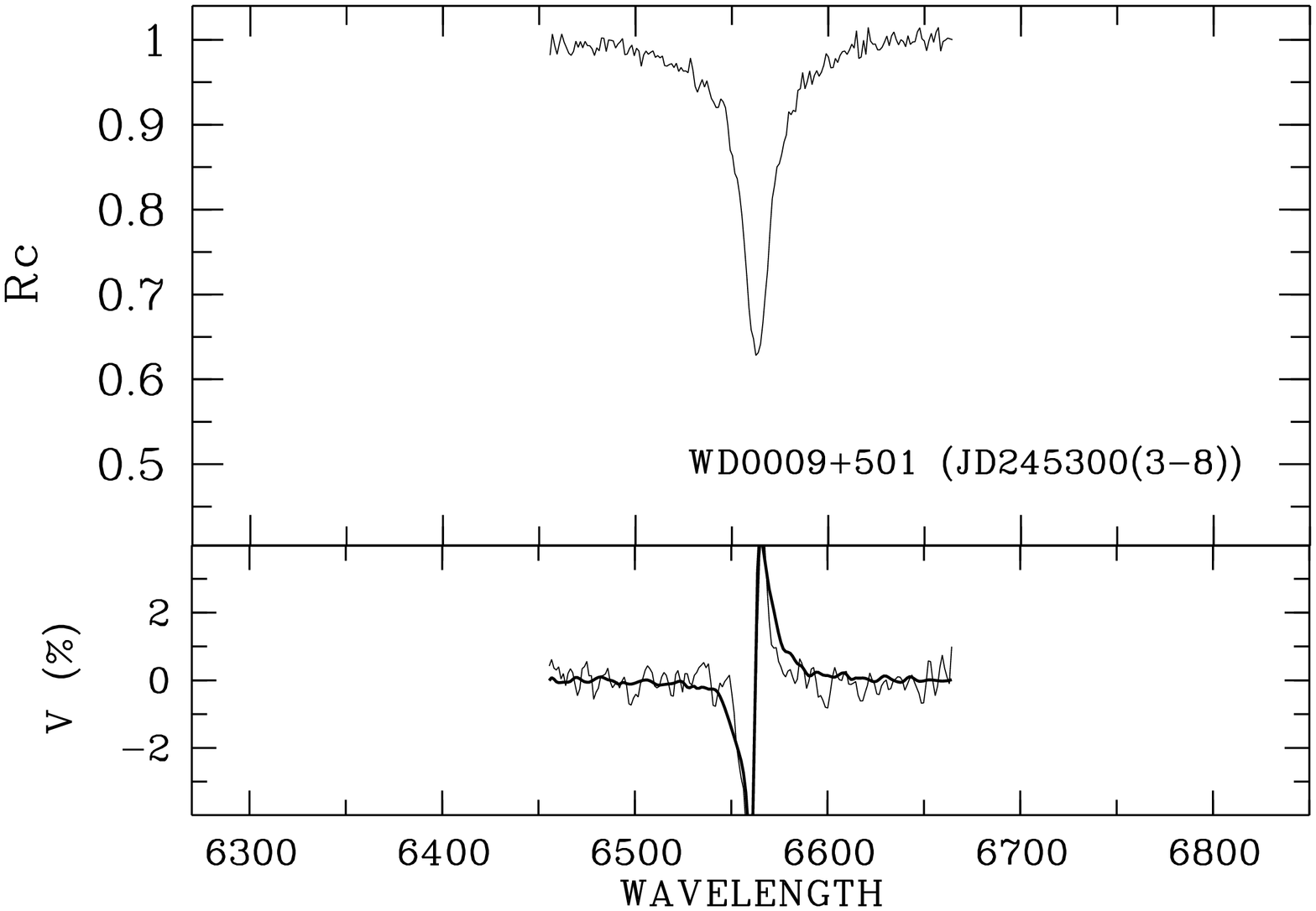}
\includegraphics[width=4.03cm, height=9.2cm, angle=270]{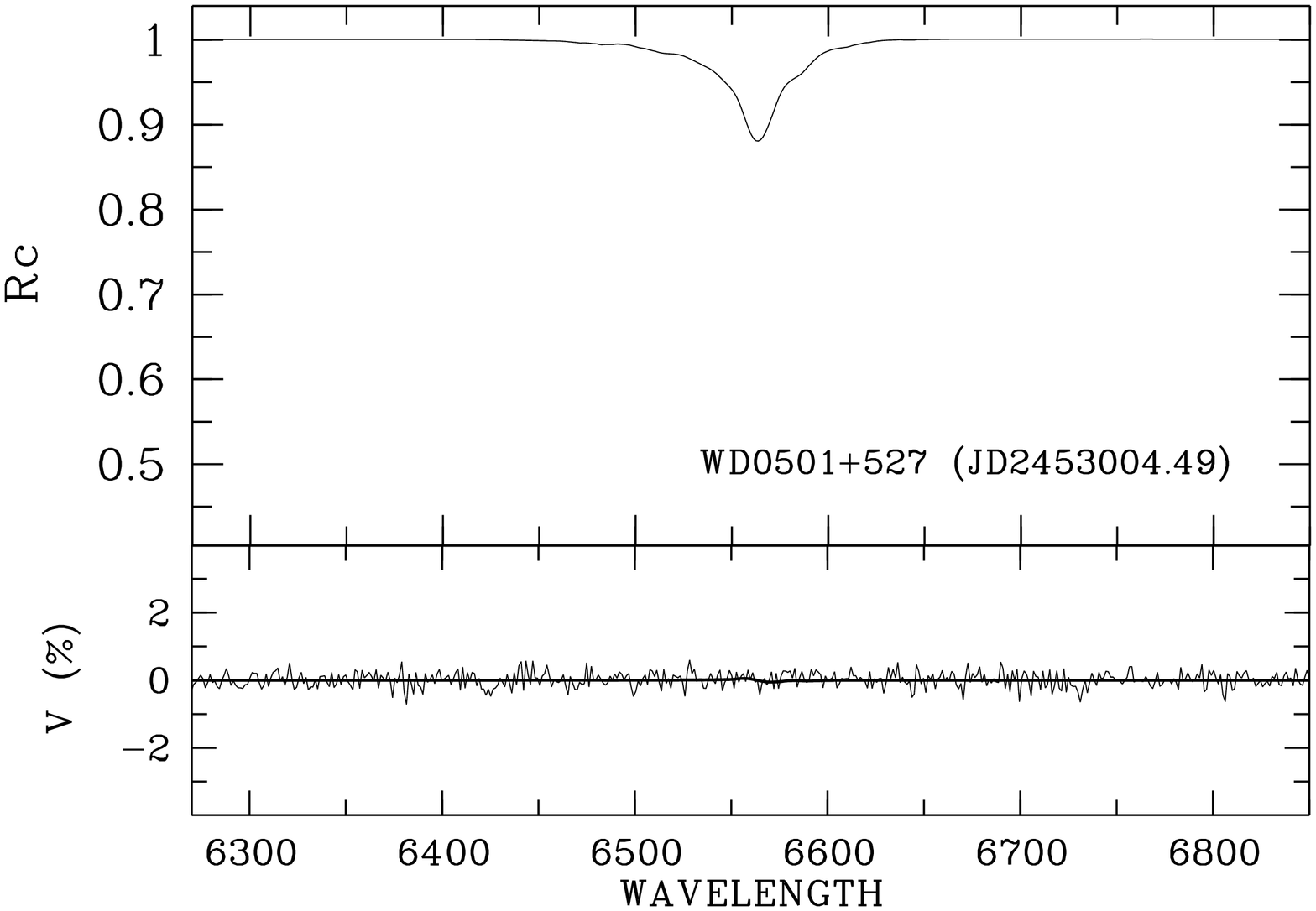}
\includegraphics[width=4.03cm, height=9.2cm, angle=270]{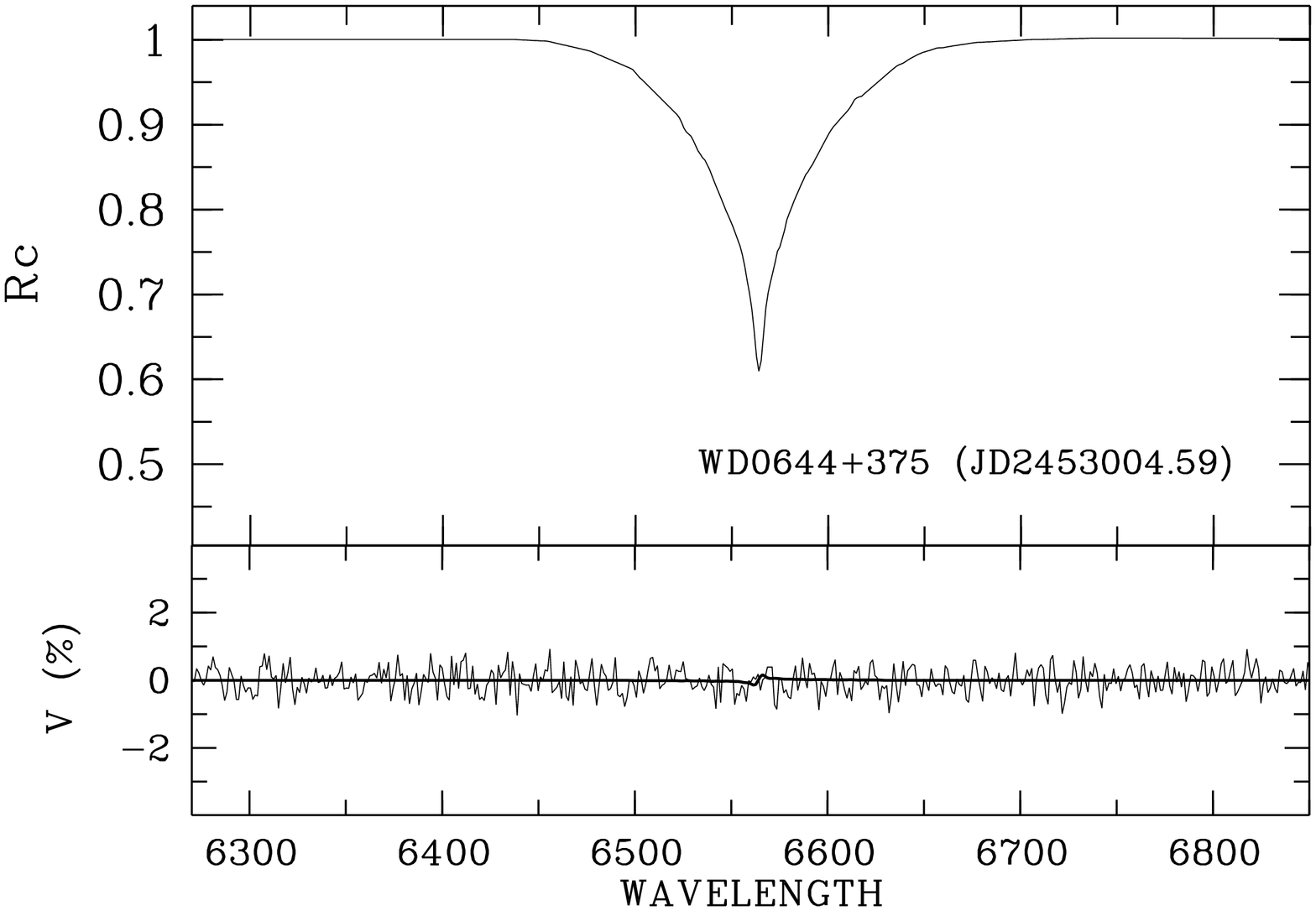}
\includegraphics[width=4.03cm, height=9.2cm, angle=270]{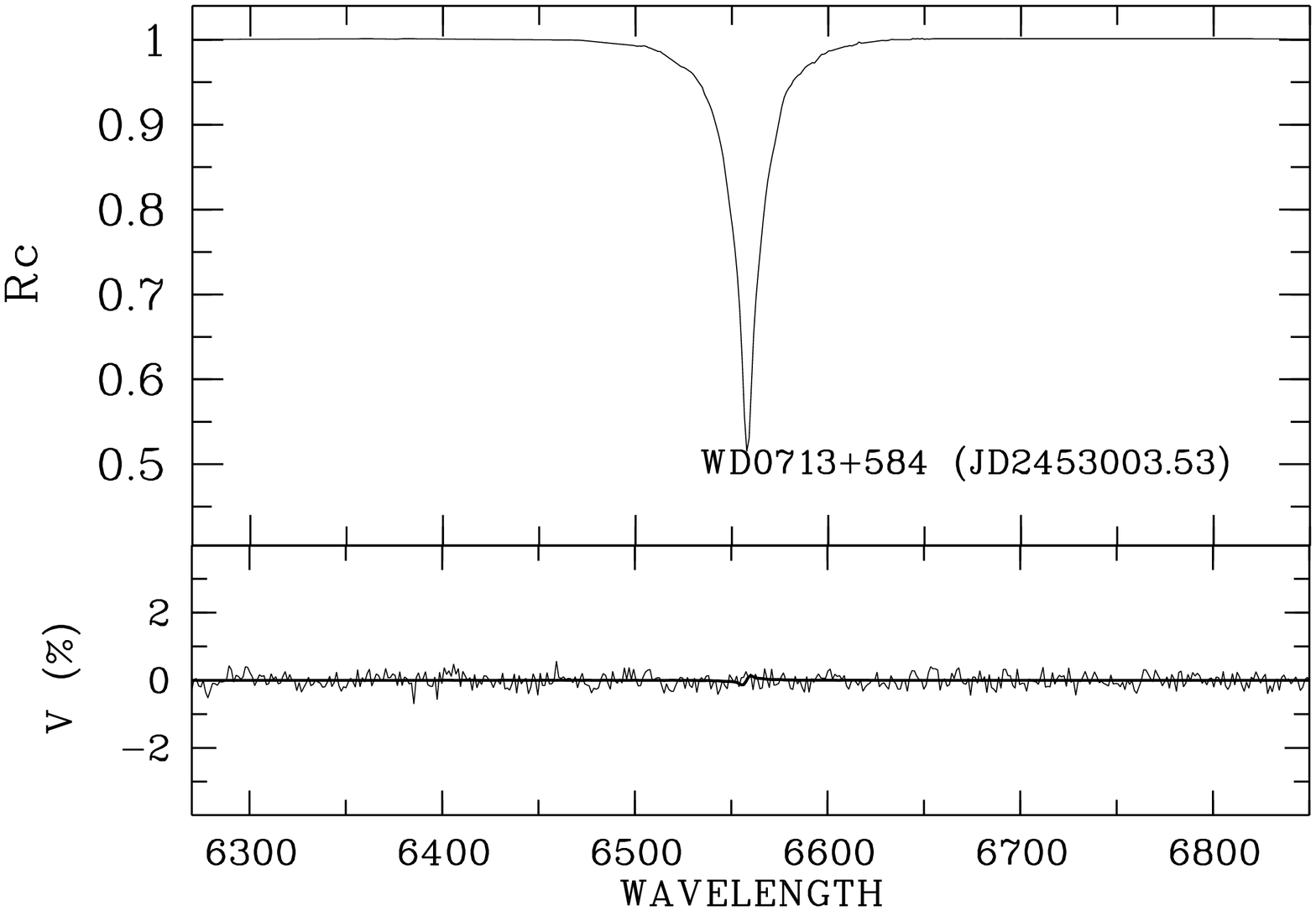}
\end{figure}
\begin{figure}
\centering
\includegraphics[width=4.03cm, height=9.2cm, angle=270]{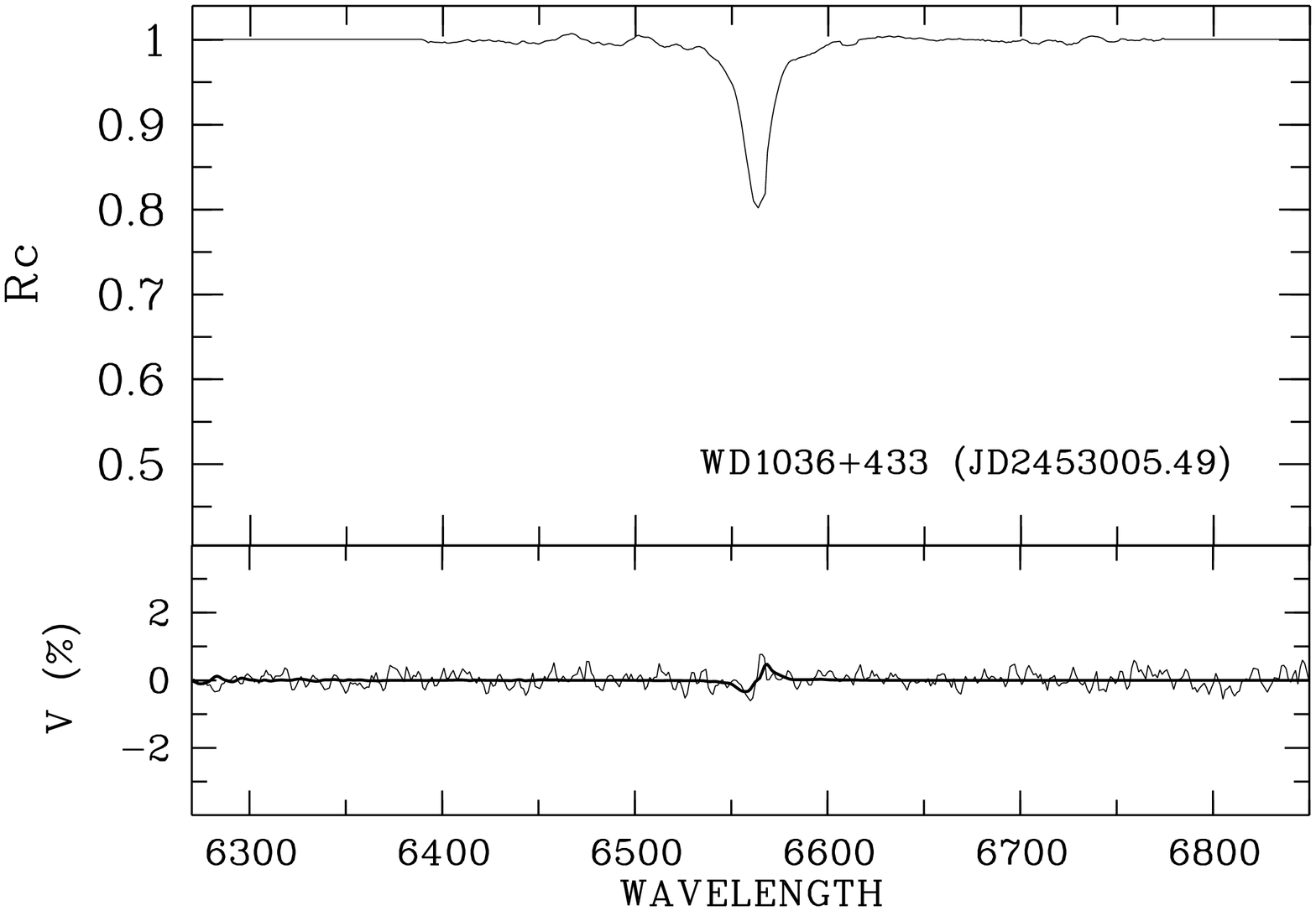}
\includegraphics[width=4.03cm, height=9.2cm, angle=270]{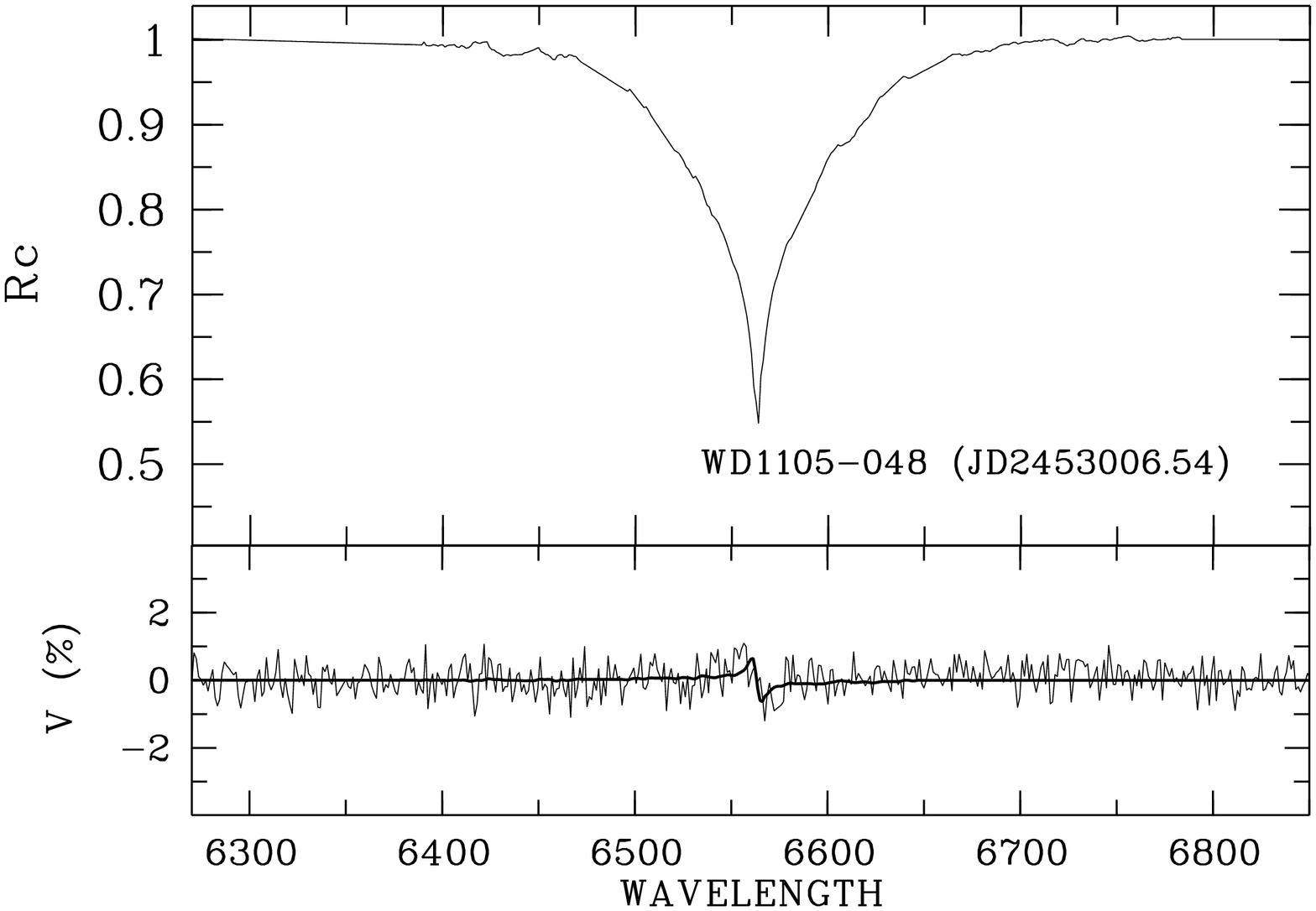}
\includegraphics[width=4.03cm, height=9.2cm, angle=270]{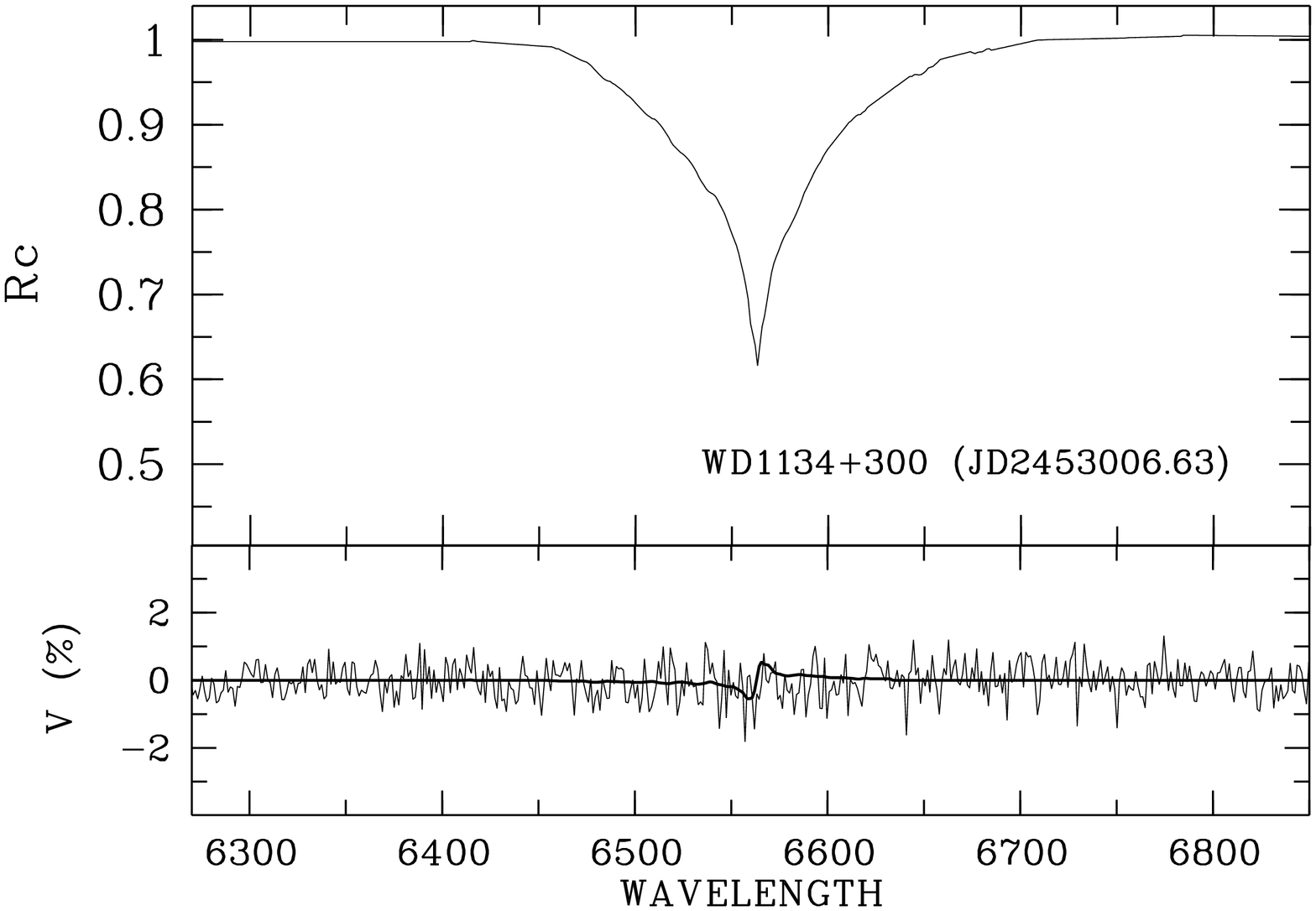}
\includegraphics[width=4.03cm, height=9.2cm, angle=270]{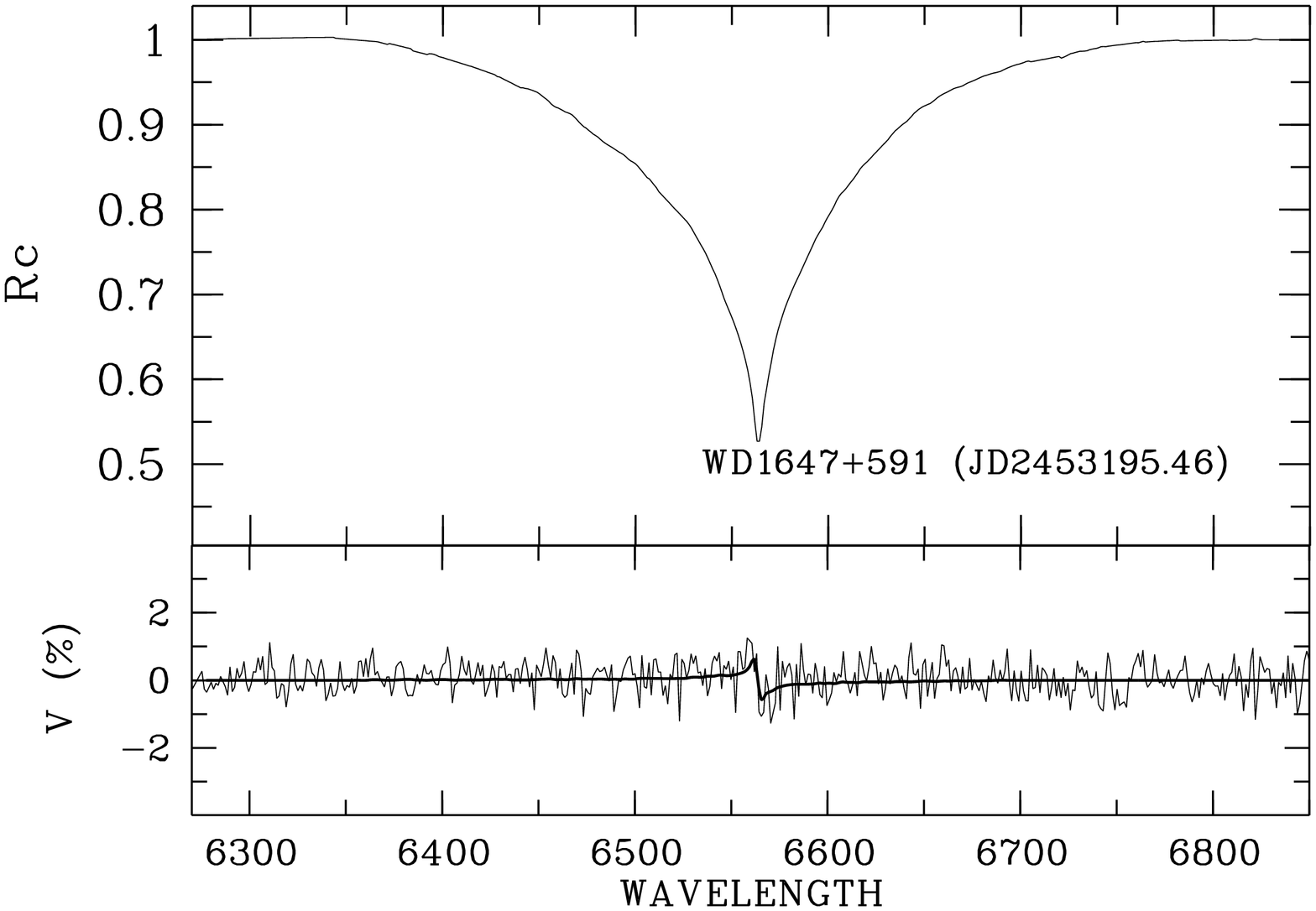}
\caption{Stokes $I$ and $V$ spectra of observed WDs at the H$\alpha$
core. The weak-field fit of the Stokes $V$ spectra is illustrated by
thick lines.}
\end{figure}
%%%%%%%%%%%%%%%%%%%%%%%%%%%%%%%%%%%%%%%%%%%%%%%%%%%%%%%%%%%%%%%%%%%%%

\end{document}